# Limit of Temporal Resolution in Atomic Force Microscopy: How fast can we image with atomically-engineered tips while preserving picometer range spatial resolution?

Omur E. Dagdeviren[1,*]

[1] Department of Physics, McGill University, Montréal, Québec, Canada, H3A 2T8

*Corresponding author's email: omur.dagdeviren@mcgill.ca

**Abstract**

With recent advances in dynamic scanning probe microscopy techniques, it is now a routine to image the sub-molecular structure of molecules with atomically-engineered tips which are prepared via controlled modification of the tip termination and are chemically well-defined. The enhanced spatial resolution is possible as atomically-engineered tips can preserve their integrity in the repulsive interaction regime. Although the mechanism of improved spatial resolution has been investigated both experimentally and theoretically, the ultimate temporal resolution while preserving picometer scale spatial resolution still remains an open question. Here, we computationally analyze the temporal resolution of atomic force microscopy imaging with atomically-engineered tips. Our computational results reveal that non-metal terminated tips, e.g. oxygen-terminated copper, are well-suited for enhanced temporal resolution up to video rate imaging velocities while preserving picometer range spatial resolution. Contrarily, the highest-attainable spatial resolution of atomically-engineered tips with low-stiffness, e.g. CO-terminated, deteriorate with increasing imaging velocity. Our results reveal that when atomically-engineered tips terminated with molecules are in use, imaging velocities in the order of nanometers per second at most are inevitable even for atomically flat surfaces to retain atomic resolution and avoid slip-stick motion. In addition to shedding light on the temporal resolution of atomic force microscopy imaging with atomically-engineered tips, our numerical results provide an outlook to the scalability of atom-by-atom fabrication using scanning probe microscopy techniques.

**Introduction**

Dynamic atomic force microscopy (AFM) is an analytical surface characterization tool where a sharp probe tip is mounted to the end of an oscillating probe and serves as a sensing element to disclose surface properties with picometer and pico-Newton resolution [1-3]. In recent years, imaging the sub-molecular structure of molecules became popular with the advent of AFM and related techniques [4, 5]. The termination of the tip apex is modified on purpose, i.e. atomically engineered, either by picking up a molecule or a chemically identified atom [4-9]. One of the most widespread ways to engineer the tip apex is to pick a CO molecule while imaging with other molecules has been demonstrated [4, 6, 7]. Also, non-metal terminated tips such as oxygen, chlorine, or bromine terminated tips are alternative to tips terminated with molecules [5, 8, 9]. Due to their inert nature, atomically-engineered tips can preserve their integrity even in the repulsive tip-sample interaction regime [4, 5]. To understand a material's properties as a function of both their structural and chemical environment as well as its responses to external stimulations in ambient or liquid environments, video-rate AFM has been developed but is currently limited to a lateral resolution of the order of ≈10 nm [10-12]. In contrast, spatial characterization at atomic length scales is common practice under ultra-high vacuum conditions; however, video-rate imaging is achieved only in scanning tunneling microscopy mode and thus restricted to electronic properties [13]. In this manuscript, we computationally explore the prospects of video rate AFM imaging with atomically-engineered tips, while preserving picometer range spatial resolution.

Our numerical results reveal that the temporal resolution depends on the tip-sample interaction kinetics which is dictated by the tip termination, tip-sample interaction force, and imaging velocity. The structural deformation of the tip-apex, i.e. the closest atom to the surface, upsurges with increasing imaging speeds and with decreasing stiffness of the tip apex. Our computational results disclose that non-metal terminated tips such as oxygen-terminated copper tips are in principle can display temporal resolution up to video rate imaging velocities, while preserving picometer range spatial resolution. However, tips with lower stiffness such as molecule-terminated tips are obliged to slower imaging velocities that result in orders of magnitude longer image acquisition times. In addition to systematically examining the limits of temporal resolution in high-resolution AFM imaging with atomically-engineered tips, our numerical analysis also provides an outlook of the scalability of atom-by-atom fabrication with the dynamic scanning probe techniques and possible pathways for enhanced capability.

**Computational Methods**

We used the pioneering work of Prandtl and Tomlinson (PT) to explain the interaction of atomically sharp probe tip with the surface [14-24]. The two-dimensional Prandtl-Tomlinson model has been implemented successfully to elucidate the interaction of atomically sharp probe tip with flat surfaces, atomic steps, and alkali halide surfaces, and the most recently to reveal the effect of surface disorder, load, and sliding velocity on friction at small length scales [15, 20, 25-28]. Also, the interaction of a single molecule attached to the end of a probe tip can be explained with two-dimensional PT model [7, 14]. In this journal article, we apply the two-dimensional PT model to investigate the tip-sample interaction kinetics. As Figure 1 summarizes, the change in the tip termination induces a change in the stiffness of the tip-apex of an atomically-engineered tip which ultimately changes the tip-sample interaction kinetics and dictates the limits of spatial and temporal resolution of AFM imaging.

To disclose the tip-sample interaction kinetics of atomically-engineered tips, the first step is to model the *local* tip-sample interaction. We used 6-12 Lenard-Jones interaction potential for our calculations [29]:

$$U_{L-J} = 4\epsilon[(\sigma/r)^{12} - (\sigma/r)^6] \qquad (1)$$

In equation 1, $r$ denotes the distance between the centers of two atoms, $\epsilon$ is the depth of potential well and $\sigma$ is the distance at which the potential well vanishes ($\sigma = d/1.12$, where $d$ is the hard sphere diameter of the atom). We implemented the periodic boundary conditions along lateral directions to a simulation cell with a cross-section of 33×33 atoms and a 5-atom thick. The fast scan direction is along <110> direction of the simple cubic lattice [30]. We used typical parameters for metal atoms ($\epsilon$ = 415 meV, $d$ = 2.6 Å, Ref. [31])

The next step is to model the probe tip. As Figure 1 shows, the atomically sharp scanned probe is modeled as a single atom that is connected to the macroscopic body of the probe tip with an elastic spring ($c_x, c_y, c_z$ spring constants along $x, y$, and $z$ directions). The stiffness of the spring change with the termination of the atomically-engineered tip and is different than the cantilevers used in scanning probe microscopy experiments [7, 15, 26, 32].

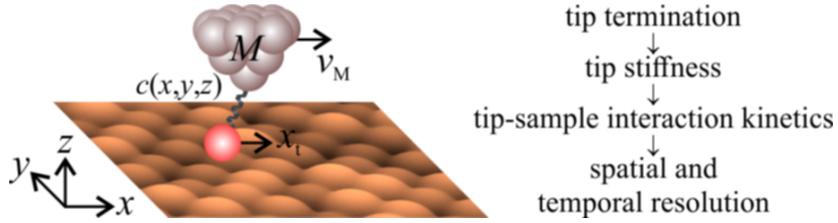

**Figure 1:** The summary of computational methods. We used two-dimensional Prandtl-Tomlinson model to investigate the kinetic interaction of the atomically-sharp probe tip with the surface as a function of tip stiffness, normal load, and imaging velocity. The tip apex is connected to the microscope body, *M*, with an elastic spring constant of $c(x, y, z)$. The body of the microscope moves along fast scan direction with the velocity $v_M$. The position of the tip apex ($x_t, y_t, z_t$) during the relative motion is determined by the tip-sample interaction kinetics. The tip-sample potential, *U,* is the energy landscape that the tip interacts with the surface.

We calculated the potential energy landscape in the vacuum up to 6.5 Ångströms with respect to the lattice position of the top layer with 2.5 picometer steps for an area of 2.6 nm × 2.6 nm in the center of the calculation slab. The distance between the tip and the sample is modulated to keep the force constant in most scanning probe microscopy experiments [2, 3]. The vertical position of the tip can be expressed as:

$$\underbrace{c_z(z_M - z_{t,0})}_{F_z = constant} = \left.\frac{\partial U_{int}}{\partial z_t}\right|_{x_t, y_t} \qquad (2)$$

In equation 2, $z_M$ is the position of the macroscopic body of the tip along the vertical direction. The stable position of the single-atom asperity along $z$-direction, $z_{t,0}$, for a fixed lateral coordinate ($x_t, y_t$) is calculated by equating the force acting on the spring-mass system along vertical direction, $F_z$ (the left-hand side of Equation 2), to the vertical component of the tip-sample interaction force for the fixed lateral coordinate (the right-hand side), which is calculated as the negative gradient of the total tip-sample interaction potential, $U_{int}$. The vertical height profile, the surface topography, is calculated for an area of 2.6 nm × 2.6 nm in the center of the calculation slab. With the calculation of equilibrium position of the apex atom along the $z$-direction, the three-dimensional tip-sample interaction potential can be reduced to a two-dimensional interaction potential. The following two-dimensional system of coupled second-order differential equations is solved to calculate the motion of the atomically sharp tip along lateral directions ($x$ and $y$ directions):

$$m_x \ddot{x}_t = c_x(x_M - x_{t,0}) - \frac{\partial U(x_t, y_t)}{\partial x_t} - \gamma_x \dot{x}_t$$

$$m_y \ddot{y}_t = c_y(y_M - y_{t,0}) - \frac{\partial U(x_t, y_t)}{\partial y_t} - \gamma_y \dot{y}_t$$

(3)

Equation 3 is solved by using ode45 function in MATLAB and restricted the relative error of the numerical solution to $10^{-10}$ [33]. In equation 3, $m_x$ and $m_y$, $10^{-8}$ kg, are effective masses of the system [20, 25]. Coordinates of the tip-apex along lateral directions are expressed as $x_t$ and $y_t$, and time derivatives of lateral positions present velocity ($\dot{x}_t, \dot{y}_t$) and acceleration ($\ddot{x}_t, \ddot{y}_t$) of the atomically sharp probe tip. Negative gradient of tip-sample interaction potential (*U)*, gives the lateral force components of the interaction potential [34].

When the tip traces the surface, the kinetic energy [35] of the tip will be dissipated [22, 36-38]. Different mechanisms such as electronic [36, 37, 39-42], electromagnetic [43], van der Walls friction [44, 45], and phononic [46-49] have been proposed for the energy dissipation due to kinetic tip-sample interaction. We included the effect of energy dissipation due to the motion of the tip with the damping coefficient term, γ, in equation 3. Figure 2 summarizes numerical solution of equation 3 in for average slip length as a function of the damping coefficient. As Figure 2 discloses, under the critical damping condition ($2\sqrt{c \times M}$, $c$ and $M$ are spring constant and effective mass of the system), the average deformation is at the order of the unperturbed lattice constant (2.6 Å) of the model system. Tip oscillations are evident for the under-damped case. For the strongly over-damped case, however, the tip sticks to a lattice site before slipping multiple lattice constants, i.e. the intrinsic contribution of the surface to the motion of the tip disappears. We used the critical damping coefficient in our calculations to eliminate tip oscillations and preserve contributions of the sample to the motion of the tip with our choice of damping coefficient.

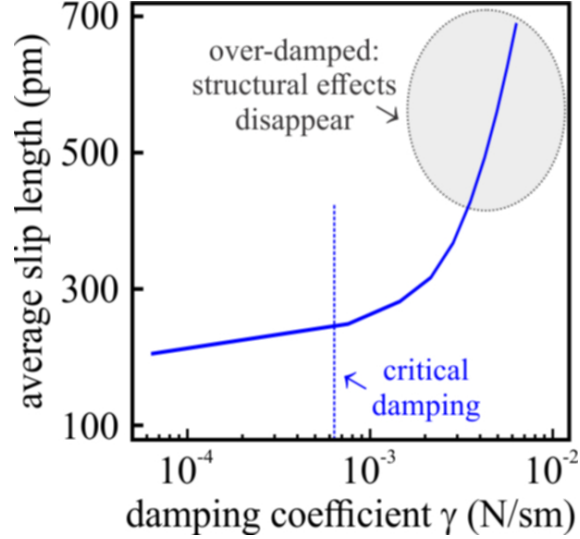

**Figure 2:** The choice of damping coefficient, γ, to model the kinetic energy dissipation of the tip due to interaction with the surface. When the system is under-damped, strong tip oscillations are induced while the contribution of the surface is diminished for the over-damped case. For these reasons, we used critical damping condition to eliminate the tip oscillations while preserving the effect of the surface on the kinetic tip-sample interaction. We solved equation 3 for a model tip with the stiffness of 10 N/m and the effective mass, M, equal to $10^{-8}$ kg for a normal load of 0.25 nN and a sliding velocity of 1,000 nm/s for calculations presented in Figure 2.

Depending on the kinetic tip-sample interaction, the tip apex can either follow the minimum energy path or slip-stick motion may be evident which impedes tracing the minimum energy trajectory and results in large structural deformations of the tip apex [27]. For this reason, the deformation length of the apex can be used as a caliber to quantify the kinetic tip-sample interaction between the tip and the sample. The total perturbation of the spring system from its equilibrium position, i.e. the deformation length, is calculated by subtracting the position of the tip apex $(x_t, y_t)$ from the unperturbed position of the macroscopic body of the microscope $(x_M, y_M)$. At the end of the scan frame, $y_M$ is changed and a new line is calculated. To simulate the movement of the scanning force microscopy experiments, the tip starts with zero velocity at the left border of the scan area ($\dot{x}_t = 0$, $y_t = 0$) with relaxed springs ($x_M = 0$, $y_M = 0$). The transient part of the solution disappears within the first 5-6 Ångströms for initial conditions defined for our calculations (1 nm/s $\leq \dot{x}_M \leq$ 10,000 nm/s). We excluded the transient part of the solution in our statistical analyses.

**Results and Discussions**

We calculated the root-mean-square of the deformation length of the atomically sharp tip as a function of imaging velocity, the tip stiffness, and normal load for an area of 2.6 nm × 2.6 nm at the center of the calculation slab. As Figure 3 reveals, three distinct regions can be identified. In region I, the deformation length of the atomically sharp probe is less than 0.1 pm, which is significantly smaller than the highest spatial resolution that can be achieved with scanning probe techniques [34]. As highlighted in region I, the imaging velocity is less than a few hundred nanometers per second for atomically sharp tips with stiffness values that are in the order of hundreds of nano-Newtons per meter. Although the atomically-sharp probe tip sustains its integrity with sub-picometer deformation in region I, the spatial resolution is constraint due to instrumental limitations such as readout noise, mechanical stability [50]. In region II, the spatial resolution is also preserved for high-resolution imaging. The deformation length of the tip apex is between 0.1 pm to 10.0 pm which fits in the range of atomic-resolution images, and video rate imaging velocities for atomic resolution images (1,000 nm/s, roughly 400 lines per/second) are attainable. Contrarily in region III, the tip reveals deformations comparable to the lattice constant and the slip-stick motion is evident. Also, the slip length inflates with decreasing tip stiffness and slow imaging velocities in the order of sub-nanometer per second are inevitable to preserve the spatial resolution.

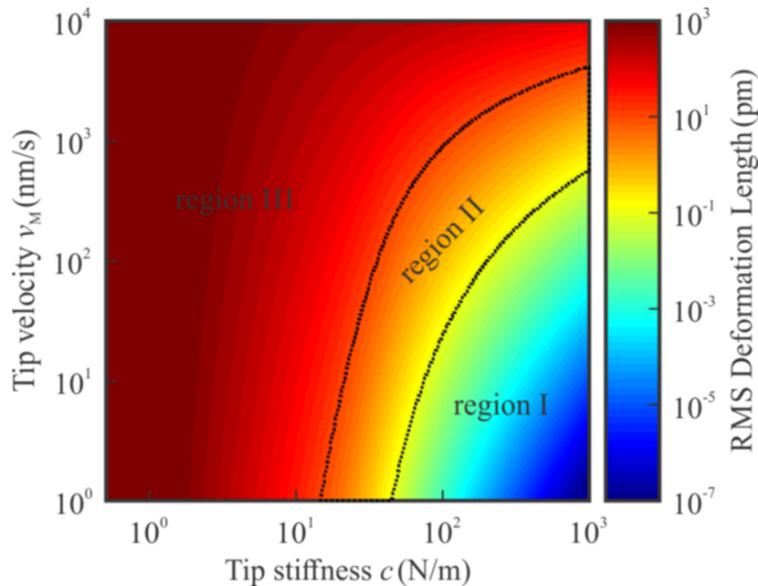

**Figure 3:** The root-mean-square of the deformation length of atomically sharp tip as a function of imaging velocity and tip stiffness. Three distinct regions are evident. The spatial resolution is restraint due to instrumental limitations in region I in which the deformation length is less than 0.1 picometer. In region II, stiffness of the atomically sharp tip is in the order of tens of nano-Newtons and picometer range spatial resolution is preserved up to video rate imaging velocities for nanometer size scans. The slip length limits the spatial resolution in region III, i.e. slow imaging speeds are inevitable with the upsurge of the deformation length. The total vertical force acting on the tip is 200 pico-Newton for calculations presented in Figure 3.

To elucidate the kinetic interaction of the atomically sharp probe tip and the surface, we investigated the deformation length as a function of normal load, tip stiffness, and the imaging velocity. As Figure 4 reveals, we first explored the interaction of a tip with 5 N/m stiffness, which is a comparable to atomically-engineered tips with molecules [6, 7]. As Figure 4 a and b reveal, the interaction of the atomically sharp tip displays slip-stick motion (for details see inset in Figure 4a) which is consistent with the experiments conducted with atomically engineered tips terminated with molecules [7]. Also, Figure 4c displays, multiple slip-stick events are evident with increasing load for imaging speeds in the order of 1,000 nm/s. Besides, we examined the kinetic tip-sample interaction of a tip with 50 N/m which has a similar the stiffness of an oxygen-terminated copper tip [5, 9]. As highlighted by figure 4 d-f, slip-stick events are not evident for the stiffer tip (for details see inset in Figure 4d). Even though the deformation length increases with the load, the root-mean-square of the deformation length is in the order of 10 picometers even for repulsive forces acting on the tip and the smooth tip motion is preserved with the absence of slip-stick motion.

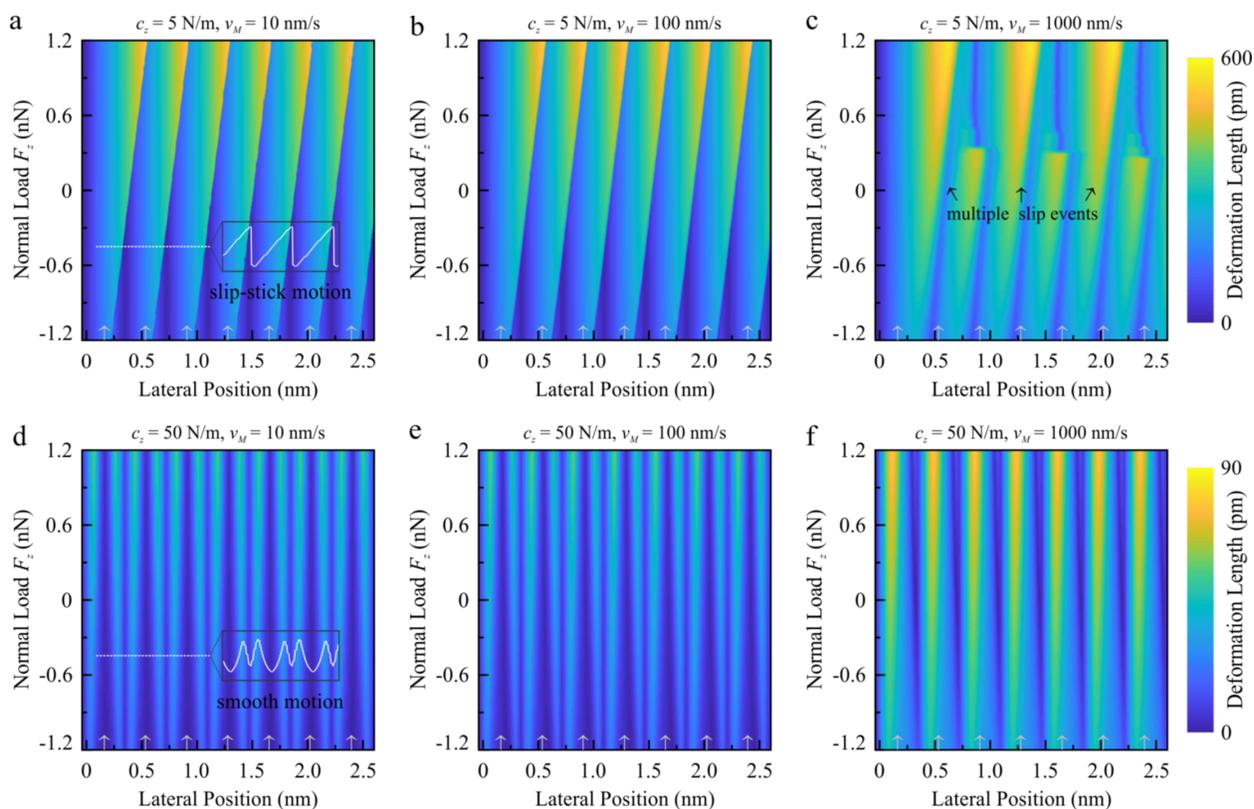

**Figure 4:** Calculation of the deformation length as a function of tip stiffness, normal load, and imaging velocity for two different tip stiffness. (a-c) The stiffness values that are comparable to the stiffness of a molecule-terminated tip discloses slip-stick motion (e.g. profile of the dashed line as the inset in a), which is consistent with experimental results. With increasing stiffness (d-f), e.g. oxygen-terminated copper tip, the slip-stick motion is eliminated, e.g. profile of the dashed line as the inset in (d). White arrows show the center of surface atoms (a-f).

Imaging velocities in the order of sub-nanometer per second are inevitable to achieve picometer spatial resolution 'soft' atomically-engineered tips, i.e. tips terminated with a molecule. Otherwise, the spatial resolution attenuates due to slip-stick motion. As our numerical results demonstrate that with increasing imaging velocity, 'stiffer' atomically-engineered (e.g. oxygen-terminated copper) tips favors preservation of the spatial resolution. As the slip-stick motion is eliminated with stiff atomically-engineered tips, the atomically sharp tip can follow the lowest energy path [27]. For this reason, it is important to note that the tip deforms along both lateral directions. The deformation of the tip-apex along the slow-scan axis, i.e. the direction perpendicular to $v_M$, may alter the interpretation of multi-dimensional force measurements or measured forces [51] and measured energy barriers of manipulation experiments [52].

Our computational analysis does not take into account the additional noise sources and limitations of measurement electronics which may further scale down the maximum attainable imaging velocities. Also, the deformation of the surface or structures imaged (e.g. molecules), can constraint imaging velocity. Although we use a simple model, our results reveal that the imaging chemical reactions in real time with conventional atomic force microscopy imaging have challenges with the available set of atomically-engineered tips and due to instrumental constraints. Also, our results disclose that scaling atom-by-atom fabrication with atomically engineered tips has limitations if the atomic structure of the scanning probe is important.

**Summary**

We conducted numerical analysis to investigate the kinetics of tip-sample interaction of atomically-engineered tips to reveal the limits of temporal resolution while preserving picometer range spatial resolution by using two-dimensional Prandtl-Tomlinson model. Our numerical results disclose that the maximum temporal resolution of AFM imaging with atomically-engineered tips is dictated by the stiffness of the tip apex. The structural deformation of the apex inflates with decreasing stiffness of the tip and increasing imaging speed and normal load. Our calculations show that '*stiff*' atomically-engineered tips such as oxygen-terminated copper tips can withstand video rate imaging velocities for atomic-resolution images. Contrarily, tips with lower stiffness, e.g. molecule-terminated tips, are confined to slower imaging velocities in the order of nanometers per second at most to preserve picometer-range spatial resolution. Besides, our results highlight that the scalability of atom-by-atom fabrication with the dynamic probe techniques has major challenges, if the atomic structure of the probe is important.

**Acknowledgements**

I would like to thank Prof. Peter Grütter and Dr. Yoichi Miyahara for fruitful discussions. Financial support from The Natural Sciences and Engineering Research Council of Canada and Le Fonds de Recherche du Québec - Nature et Technologies are gratefully acknowledged.

**References**

1. T.R. Albrecht, P. Grutter, D. Horne, and D. Rugar, Frequency modulation detection using high-Q cantilevers for enhanced force microscope sensitivity. Journal of Applied Physics **69**, (1991).

2. R. Garcia, *Amplitude Modulation Atomic Force Microscopy*. 2010, Singapore: Wiley-VCH.

3. F.J. Giessibl, Advances in atomic force microscopy. Reviews of Modern Physics **75**, 949-983 (2003).

4. L. Gross, F. Mohn, N. Moll, P. Liljeroth, and G. Meyer, The Chemical Structure of a Molecule Resolved by Atomic Force Microscopy. Science **325**, 1110-1114 (2009).

5. H. Mönig, D.R. Hermoso, O. Díaz Arado, M. Todorović, A. Timmer, S. Schüer, G. Langewisch, R. Pérez, and H. Fuchs, Submolecular Imaging by Noncontact Atomic Force Microscopy with an Oxygen Atom Rigidly Connected to a Metallic Probe. ACS Nano **10**, 1201-1209 (2016).

6. A.J. Weymouth, T. Hofmann, and F.J. Giessibl, Quantifying Molecular Stiffness and Interaction with Lateral Force Microscopy. Science **343**, 1120 (2014).

7. R. Pawlak, W. Ouyang, A.E. Filippov, L. Kalikhman-Razvozov, S. Kawai, T. Glatzel, E. Gnecco, A. Baratoff, Q. Zheng, O. Hod, M. Urbakh, and E. Meyer, Single-Molecule Tribology: Force Microscopy Manipulation of a Porphyrin Derivative on a Copper Surface. ACS Nano **10**, 713-722 (2016).

8. S. Kawai, A.S. Foster, T. Björkman, S. Nowakowska, J. Björk, F.F. Canova, L.H. Gade, T.A. Jung, and E. Meyer, Van der Waals interactions and the limits of isolated atom models at interfaces. Nature Communications **7**, 11559 (2016).

9. F. Mohn, B. Schuler, L. Gross, and G. Meyer, Different tips for high-resolution atomic force microscopy and scanning tunneling microscopy of single molecules. Applied Physics Letters **102**, 073109 (2013).

10. T. Ando, N. Kodera, E. Takai, D. Maruyama, K. Saito, and A. Toda, A high-speed atomic force microscope for studying biological macromolecules. Proceedings of the National Academy of Sciences **98**, 12468 (2001).

11. T. Ando, T. Uchihashi, and T. Fukuma, High-speed atomic force microscopy for nano-visualization of dynamic biomolecular processes. Progress in Surface Science **83**, 337-437 (2008).

12. P.K. Hansma, G. Schitter, G.E. Fantner, and C. Prater, High-Speed Atomic Force Microscopy. Science **314**, 601 (2006).

13. F. Besenbacher, E. Lægsgaard, and I. Stensgaard, Fast-scanning STM studies. Materials Today **8**, 26-30 (2005).

14. U.D. Schwarz and H. Hölscher, Exploring and Explaining Friction with the Prandtl–Tomlinson Model. ACS Nano **10**, 38-41 (2016).

15. P. Steiner, R. Roth, E. Gnecco, A. Baratoff, and E. Meyer, Angular dependence of static and kinetic friction on alkali halide surfaces. Physical Review B **82**, 205417 (2010).

16. C.M. Mate, G.M. McClelland, R. Erlandsson, and S. Chiang, Atomic-scale friction of a tungsten tip on a graphite surface. Physical Review Letters **59**, 1942-1945 (1987).

17. E. Gnecco, R. Bennewitz, T. Gyalog, C. Loppacher, M. Bammerlin, E. Meyer, and H.J. Güntherodt, Velocity Dependence of Atomic Friction. Physical Review Letters **84**, 1172-1175 (2000).

18. E. Riedo, E. Gnecco, R. Bennewitz, E. Meyer, and H. Brune, Interaction Potential and Hopping Dynamics Governing Sliding Friction. Physical Review Letters **91**, 084502 (2003).

19. L. Jansen, H. Hölscher, H. Fuchs, and A. Schirmeisen, Temperature Dependence of Atomic-Scale Stick-Slip Friction. Physical Review Letters **104**, 256101 (2010).

20. H. Hölscher, D. Ebeling, and U.D. Schwarz, Friction at Atomic-Scale Surface Steps: Experiment and Theory. Physical Review Letters **101**, 246105 (2008).

21. D. Gangloff, A. Bylinskii, I. Counts, W. Jhe, and V. Vuletić, Velocity tuning of friction with two trapped atoms. Nature Physics **11**, 915 (2015).

22. C. Fusco and A. Fasolino, Velocity dependence of atomic-scale friction: A comparative study of the one- and two-dimensional Tomlinson model. Physical Review B **71**, 045413 (2005).

23. L. Prandtl, Ein Gedankenmodell zur kinetischen Theorie der festen Körper. ZAMM - Journal of Applied Mathematics and Mechanics / Zeitschrift für Angewandte Mathematik und Mechanik **8**, 85-106 (1928).

24. G.A. Tomlinson, CVI. A molecular theory of friction. The London, Edinburgh, and Dublin Philosophical Magazine and Journal of Science **7**, 905-939 (1929).

25. H. Hölscher, U.D. Schwarz, and R. Wiesendanger, Modelling of the scan process in lateral force microscopy. Surface Science **375**, 395-402 (1997).

26. P. Steiner, R. Roth, E. Gnecco, A. Baratoff, S. Maier, T. Glatzel, and E. Meyer, Two-dimensional simulation of superlubricity on NaCl and highly oriented pyrolytic graphite. Physical Review B **79**, 045414 (2009).

27. O.E. Dagdeviren, Exploring load, velocity, and surface disorder dependence of friction with one-dimensional and two-dimensional models. Nanotechnology **29**, 315704 (2018).

28. O.E. Dagdeviren, Nanotribological properties of bulk metallic glasses. Applied Surface Science **458**, 344-349 (2018).

29. J. Israelachvili, *Intermolecular and Surface Forces*. 2 ed. 1991, London: Academic Press.

30. N.W.a.M. Ashcroft, N.D., *Solid State Physics*. 1981, Philadelphia: Sounders College.

31. P.M. Agrawal, B.M. Rice, and D.L. Thompson, Predicting trends in rate parameters for self-diffusion on FCC metal surfaces. Surface Science **515**, 21-35 (2002).

32. M.Z. Baykara, O.E. Dagdeviren, T.C. Schwendemann, H. Mönig, E.I. Altman, and U.D. Schwarz, Probing three-dimensional surface force fields with atomic resolution: Measurement strategies, limitations, and artifact reduction. Beilstein Journal of Nanotechnology **3**, 637-650 (2012).

33. Mathworks, *MATLAB computing environment and programming language*. 2017, Mathworks, Natick, MA, USA.

34. R. Wiesendanger, *Scanning Probe Microscopy and Spectroscopy: Methods and Applications*. 1994: Cambridge University Press.

35. O.E. Dagdeviren, J. Götzen, H. Hölscher, E.I. Altman, and U.D. Schwarz, Robust high-resolution imaging and quantitative force measurement with tuned-oscillator atomic force microscopy. Nanotechnology **27**, 065703 (2016).

36. B. Bhushan, J.N. Israelachvili, and U. Landman, Nanotribology: friction, wear and lubrication at the atomic scale. Nature **374**, 607 (1995).

37. J.B. Sokoloff, Theory of dynamical friction between idealized sliding surfaces. Surface Science **144**, 267-272 (1984).

38. S.Y. Krylov and J.W.M. Frenken, The physics of atomic-scale friction: Basic considerations and open questions. physica status solidi (b) **251**, 711-736 (2014).

39. J.B. Sokoloff, Theory of the contribution to sliding friction from electronic excitations in the microbalance experiment. Physical Review B **52**, 5318-5322 (1995).

40. J.E. Sacco, J.B. Sokoloff, and A. Widom, Dynamical friction in sliding condensed-matter systems. Physical Review B **20**, 5071-5083 (1979).

41. V.L. Popov, Electronic contribution to sliding friction in normal and superconducting states. Journal of Experimental and Theoretical Physics Letters **69**, 558-561 (1999).

42. V.L. Popov, Electronic and phononic friction of solids at low temperatures. Tribology International **34**, 277-286 (2001).

43. V.D. Georgii, Nanotribology: experimental facts and theoretical models. Physics-Uspekhi **43**, 541 (2000).

44. B. Gotsmann, Sliding on vacuum. Nature Materials **10**, 87 (2011).

45. A.I. Volokitin, B.N.J. Persson, and H. Ueba, Enhancement of noncontact friction between closely spaced bodies by two-dimensional systems. Physical Review B **73**, 165423 (2006).

46. J.B. Sokoloff, Theory of energy dissipation in sliding crystal surfaces. Physical Review B **42**, 760-765 (1990).

47. J.B. Sokoloff, Theory of atomic level sliding friction between ideal crystal interfaces. Journal of Applied Physics **72**, 1262-1270 (1992).

48. V.L. Popov, Superslipperiness at Low Temperatures: Quantum Mechanical Aspects of Solid State Friction. Physical Review Letters **83**, 1632-1635 (1999).

49. S. Kajita, H. Washizu, and T. Ohmori, Approach of semi-infinite dynamic lattice Green's function and energy dissipation due to phonons in solid friction between commensurate surfaces. Physical Review B **82**, 115424 (2010).

50. C.J. Chen, *Introduction to Scanning Tunneling Microscopy*. 1993, New York: Oxford University Press.

51. B.J. Albers, T.C. Schwendemann, M.Z. Baykara, N. Pilet, M. Liebmann, E.I. Altman, and U.D. Schwarz, Three-dimensional imaging of short-range chemical forces with picometre resolution. Nat Nano **4**, 307-310 (2009).

52. G. Langewisch, J. Falter, H. Fuchs, and A. Schirmeisen, Forces During the Controlled Displacement of Organic Molecules. Physical Review Letters **110**, 036101 (2013).